\documentclass[aps, prl, reprint, amsmath, amssymb, showkeys, unsortedaddress]{revtex4-2}

\usepackage{graphicx}
\usepackage{dcolumn}
\usepackage{bm}
\usepackage{siunitx}
\usepackage{xcolor, soul}
\usepackage[utf8]{inputenc}

\begin{document}

    \title[]{Three-dimensional coherent Bragg imaging of rotating nanoparticles}
    
    \author{Alexander Björling}
    \email{alexander.bjorling@maxiv.lu.se}
    \affiliation{MAX IV Laboratory, Lund University, 22100 Lund, Sweden}
    
    \author{Lucas A. B. Mar\c{c}al}
    \affiliation{Synchrotron Radiation Research and NanoLund, Lund University, 22100 Lund, Sweden}
    
    \author{José Solla-Gullón}
    \affiliation{Institute of Electrochemistry, University of Alicante, 03080 Alicante, Spain}
    
    \author{Jesper Wallentin}
    \affiliation{Synchrotron Radiation Research and NanoLund, Lund University, 22100 Lund, Sweden}
    
    \author{Dina Carbone}
    \affiliation{MAX IV Laboratory, Lund University, 22100 Lund, Sweden}
    
    \author{Filipe R. N. C. Maia}
    \affiliation{Department of Cell and Molecular Biology, Uppsala University, 75124 Uppsala, Sweden}
    
    \date{\today}
    
    \begin{abstract}
        Bragg Coherent Diffraction Imaging (BCDI) is a powerful strain imaging tool, often limited by beam-induced sample instability for small particles and high power densities. Here, we devise and validate an adapted diffraction volume assembly algorithm, capable of recovering three-dimensional datasets from particles undergoing uncontrolled and unknown rotations. We apply the method to gold nanoparticles which rotate under the influence of a focused coherent X-ray beam, retrieving their three-dimensional shapes and strain fields. The results show that the sample instability problem can be overcome, enabling the use of fourth generation synchrotron sources for BCDI to their full potential.
    \end{abstract}
    
    \maketitle
    
    \section{Introduction}
    
        Bragg Coherent Diffraction Imaging (BCDI) has emerged as a promising technique for three-dimensional strain imaging of crystalline nanoparticles~\cite{Pfeifer2006,Robinson2009}. The method is based on the iterative phase-retrieval of a set of two-dimensional diffraction patterns, which together form a volume of intensities~\cite{Berenguer2013}. Phased diffraction volumes provide detailed three-dimensional maps of how strain and defects are distributed within single particles~\cite{Favre-Nicolin2010, Newton2010}. While it offers a lower resolution than electron microscopy or scanning probe methods, BCDI exploits the penetration depth and high strain sensitivity of X-rays and is therefore applicable to demanding environments such as catalysts in working reactors or cells~\cite{Ulvestad2014,Ulvestad2015,Ulvestad2016,Ulvestad2017,Liu2017,Meirer2018,Kim2018,Kim2019,Abuin2019}. As X-ray sources develop worldwide and higher coherent flux densities are made available, smaller particles can in principle be studied. Unfortunately, particle stability under the intense beam often becomes a limiting factor in practice~\cite{Bjorling2019}.
        
        \begin{figure}
            \includegraphics{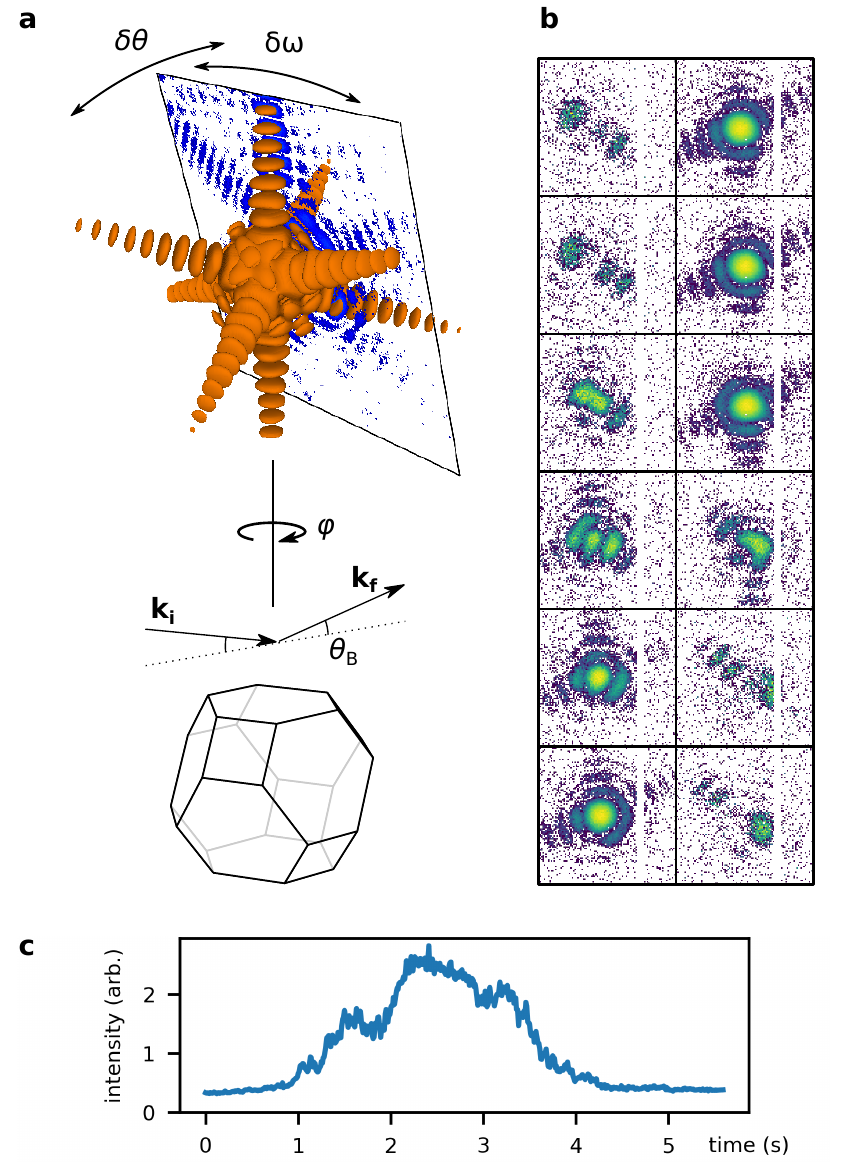}
            \caption{a) Geometrical description of the BCDI experiment. An example model particle (bottom) and its corresponding diffraction volume (top), which is sampled across the detector plane. Measured intensities are sensitive to small variations of the rocking ($\theta$) and roll ($\omega$) angles, but less sensitive to similar variations of the azimuth ($\varphi$). b) Diffraction frames recorded as a real particle rotates through the Bragg condition, with 48~ms exposure time. They correspond to cuts through the diffraction volume in (a) at unknown $\theta$ and $\omega$. The white vertical line is a gap between two detector modules. c) The total diffracted intensity of a particle undergoing an uncontrolled rocking curve, recorded at 100 Hz.}
            \label{fig:intro}
        \end{figure}

        In BCDI experiments (cf~Fig.~\ref{fig:intro}a), three-dimensional shape and strain information is encoded in a small region of reciprocal space surrounding each reciprocal lattice point $\mathbf{G}_{hkl}$. Because the reciprocal lattice rotates with the particle, and since the vector $\mathbf{G}_{hkl}$ provides a long lever arm, the entire diffraction volume can be recorded by rotating the particle through a small angular range $\delta\theta$ (see ref.~\cite{Berenguer2013} for a detailed account of the reciprocal geometry). These so-called rocking curves typically cover angles on the order of a degree~\cite{Williams2003}. Similarly, depending on the size of the detector used, small rotations $\delta\omega$ of the particle roll angle can quickly rotate the Bragg peak off the detector area, while comparable rotations of the angle $\varphi$ leave the diffraction patterns essentially unaltered. Because of this angular sensitivity, any uncontrolled rotations due to local heating or to radiation pressure~\cite{Kim2016, Liang2018} make controlled BCDI experiments of small particles very difficult. An approach to deal with small deviations from nominal rocking angles has been developed~\cite{Calvo-Almazan2019}. The method which re-formulates the phase retrieval problem is limited to small deviations from an ideal reciprocal space sampling pattern, and therefore optimized to deal with slow drifts and instrumental errors. In contrast, we present an approach that provides a solution for large uncontrolled particle rotations, where no prior knowledge of the angular trajectories across the rocking curve exists.

        We have observed entirely uncontrolled rotations of 60~nm truncated-octahedral gold nanoparticles under the focused beam at the NanoMAX beamline of the MAX IV Laboratory, as exemplified in Fig.~\ref{fig:intro}b. The 10 keV beam delivers around $2\times10^{10}$ coherent photons per second in a spot of around \SI{100}{nm}$\times$\SI{100}{nm}~\cite{Bjorling2020}. At this power density, these particles tend to rotate in and out of the diffraction condition on a timescale of seconds, completing inadvertent rocking curves around the selected reciprocal lattice point $\mathbf{G}_{111}$ in the process. As shown in Fig.~\ref{fig:intro}c, the intensity of the Bragg peak typically varies continuously but seemingly randomly between frames. Despite this, we here present a method capable of assembling data from unknown trajectories of $\theta$ and $\omega$ into coherent diffraction volumes.
        
        The results obtained demonstrate that reliable three-dimensional shape and strain images can be reconstructed from these data, assuming that the diffraction volume is sufficiently oversampled along the rocking curve. By taking advantage of beam-induced particle instability and the ensuing rotations, turning them into a resource for a more efficient data acquisition without the need for rotation motors or energy scans, we thereby show that BCDI is possible without control or even knowledge of the rocking and roll angles. This represents a substantial advance in the field of single particle inverse microscopy, opening new avenues for the exploitation of the increased coherence properties of X-ray sources.
        
    \section{Diffraction volume assembly}
    
        The present problem of assembling slices from unknown positions in a diffraction volume has a counterpart in the forward three-dimensional diffraction imaging performed at X-ray Free-Electron Lasers (XFEL:s)~\cite{Loh2010, Ekeberg2015}. In the forward geometry, all slices intersect the reciprocal space origin at three unknown angles. In the Bragg case, the slices do not intersect but can be considered virtually parallel, offset by differences in $\theta$ and $\omega$. Another key difference is that in the XFEL case, each diffraction pattern is collected from a different particle in a diffract-and-destroy manner~\cite{Chapman2006}, and assembly assumes that the original particles were identical. In the present case, a whole series of diffraction patterns such as that in Fig.~\ref{fig:intro}b is collected from one single particle.
        
        Inspired by the Expand-Maximize-Compress (EMC) algorithm used for the single-shot XFEL technique~\cite{Loh2009}, we have developed an analogous algorithm for BCDI. In essence, a model $W$ for the diffraction intensity volume is iteratively updated based on likelihood maximization with respect to the frames and their positions in $\theta$ and $\omega$. At the same time, the diffraction volume is constrained to be compatible with a particle of limited extent. Optionally, continuity of the parameters $\theta(k)$ and $\omega(k)$ as functions of time can also be imposed. We assumed the angle $\varphi$ to be constant, as a small rotation with a similar magnitude to $\delta\theta$ or $\delta\omega$ would not significantly alter the measured diffraction pattern (cf. Fig.~\ref{fig:intro}a). Similarly, the particle is assumed not to translate during the rocking-curve acquisition, which implies that the illumination conditions in the focused beam stay the same. This assumption is justified by considering that the typical displacement following a rotation would scale with the size of the rotation times the particle radius, and be of order $\delta\theta \cdot R < 1$~nm.
        
        The model is discretized as $W_{ji}$, where $j$ indexes the rocking-curve ($\theta$) bins, and $i$ runs over the detector pixels. In a first step, the probability that the $k$:th experimental frame $K_{ki}$ corresponds to $\theta$-bin $j$ and to $\omega$-bin $l$ is calculated as a matrix $P_{jlk}$. Assuming Poisson counting statistics as the main noise contributor, these likelihoods can be expressed before normalization as $R_{jlk}$, which is conveniently calculated as its logarithm to avoid numerical instability. Taking the roll angle $\omega$ into account, the probability mass function of the Poisson distribution directly gives
        \begin{eqnarray}
            \log R_{jlk} = \sum_{i} \left[ \Omega_l(K_{ki}) \cdot \log(W_{ji}) - W_{ji} \right] .
            \label{eq:Rjlk}
        \end{eqnarray}
        The $i$ index will be omitted in the following. In Eq.~\eqref{eq:Rjlk}, $\Omega_l$ is a rotation operator which compensates for $\omega$ by shifting or rotating the image $K_k$ in the detector plane according to index $l$ before comparison with the model $W_j$. There is a fundamental difference between $\omega$ and $\theta$, in that the former can be approximately compensated within each two-dimensional diffraction pattern by redistributing intensities within the frame, whereas the latter cannot. The roll operator $\Omega_l$ can therefore be chosen as appropriate. Here we approximate the roll with a rotation along a ring in the detector plane, but for small roll angles a simple shift might suffice. For high count rates in $K_k$, iterative likelihood maximization tends to get stuck in local optima, which can be avoided using an annealing parameter $\beta$ that effectively smooths the likelihood landscape \cite{Ayyer2016}. Since the probability that a measured diffraction pattern belongs somewhere in the three-dimensional model $W$ must equal one, the normalized probabilities are
        \begin{eqnarray}
            P_{jlk} = \frac{(R_{jlk})^\beta}{\sum_{j,l}(R_{jlk})^\beta}.
            \label{eq:Pjlk}
        \end{eqnarray}
        In analogy with the original EMC algorithm~\cite{Loh2009}, the resulting likelihood maximization update rule is
        \begin{eqnarray}
            W_j' = \frac{\sum_{l,k}P_{jlk}\Omega_l(K_k)}{\sum_{l,k} P_{jlk}}.
            \label{eq:M}
        \end{eqnarray}
        
        At the beginning of the assembly process, the probabilities of a frame belonging on either side of the rocking curve center can become similar. This results in an artificial symmetry of the Bragg peak, and an X-shaped matrix $P_{jk}=\sum_l{P_{jlk}}$. In reality, the trajectory $\theta(k)$ is single-valued, and the probability distribution must therefore be centered around some single angular bin $j$ for each frame $k$. To break this symmetry, a continuity bias can be imposed. Starting with the brightest frame at index $k_\text{max}$, a Gaussian distribution of variance ${n_\sigma}^2$ elements along the $j$ index, centered on the most likely angular bin $j$ of its neighbor ($k-1$ for $k>k_\text{max}$, $k+1$ for $k<k_\text{max}$) is then multiplied into $P_{jlk}$ before normalization.
        
        In a final step the maximum extent of the particle in real space is used to constrain the volume model $W_j'$. Specifically, the particle's auto-correlation function is obtained as the Fourier transform $\mathcal{F}$ of the diffraction volume. By confining the auto-correlation volume's extent along the direction corresponding to the rocking angle $\theta$, here denoted $x$, to $2L$, the particle's maximum extent is restricted to $L$. The expression for the restricted auto-correlation function is then
        \begin{eqnarray}
            \mathcal{F}(W_j'') =
                \begin{cases}
                    \mathcal{F}(W_j'), &\text{where~}|x|\leq L\\
                    0, &\text{elsewhere.}
                \end{cases}
            \label{eq:C}
        \end{eqnarray}
        The inverse Fourier transform of Eq.~\eqref{eq:C} completes an iterative update $W_j\to W_j''$.
        
        \begin{figure}
            \includegraphics{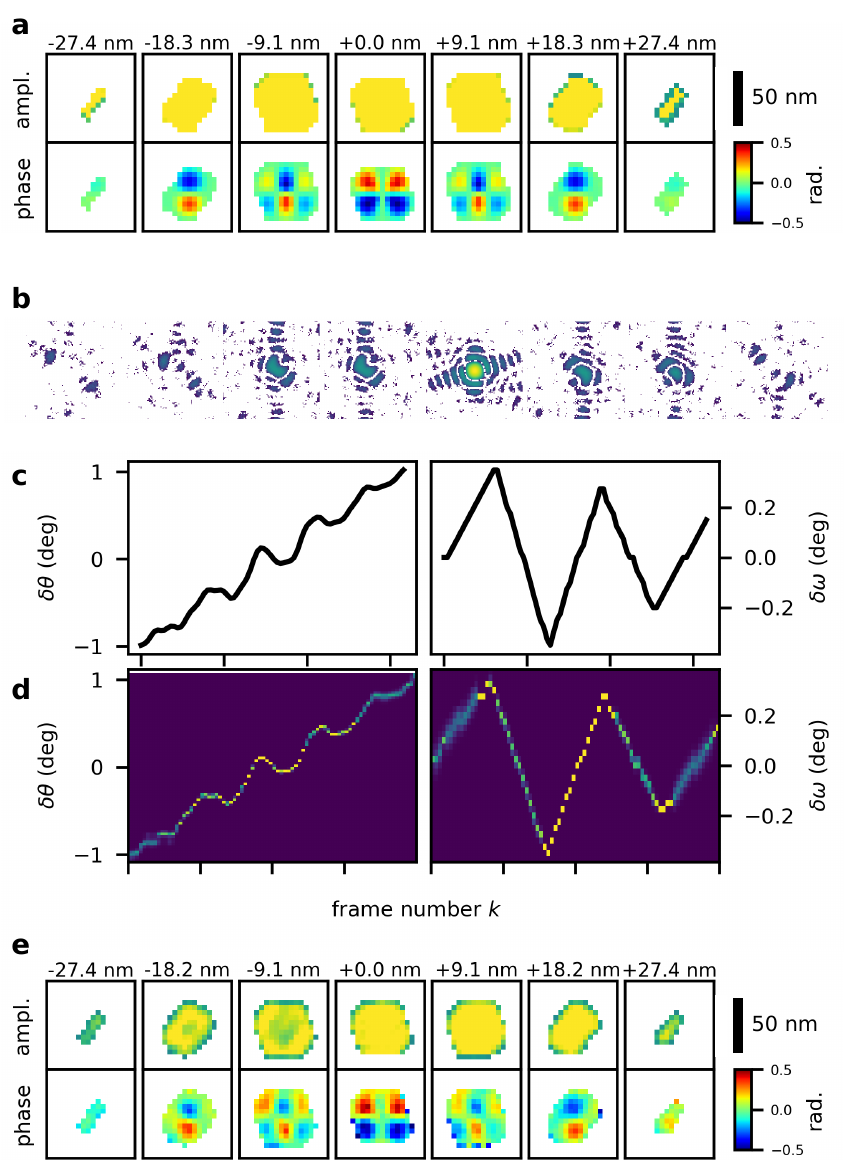}
            \caption{A numerical experiment where data is simulated for arbitrary $(\theta,\omega)$ trajectories, assembled, and phase-retrieved. a) Input model particle (cf. Fig.~\ref{fig:intro}a), described using slices of its phase and amplitude. b) Selected frames illustrating the noise level, on an arbitrary and logarithmic scale. c) The rocking and roll angles used as simulation input. d) The matrices $P_{jk}=\sum_l{P_{jlk}}$ (left) and $P_{lk}=\sum_j{P_{jlk}}$ (right) after 100 iterations of Eqns.~\eqref{eq:Rjlk}--\eqref{eq:C}. Both matrices are shown on a linear scale from 0 (navy) to 1 (yellow). e) Reconstruction of the original particle by phase retrieval of the assembled diffraction volume, as in a) above.}
            \label{fig:sim}
        \end{figure}
        
    \section{Results and validation}
    
        \begin{figure*}
            \includegraphics{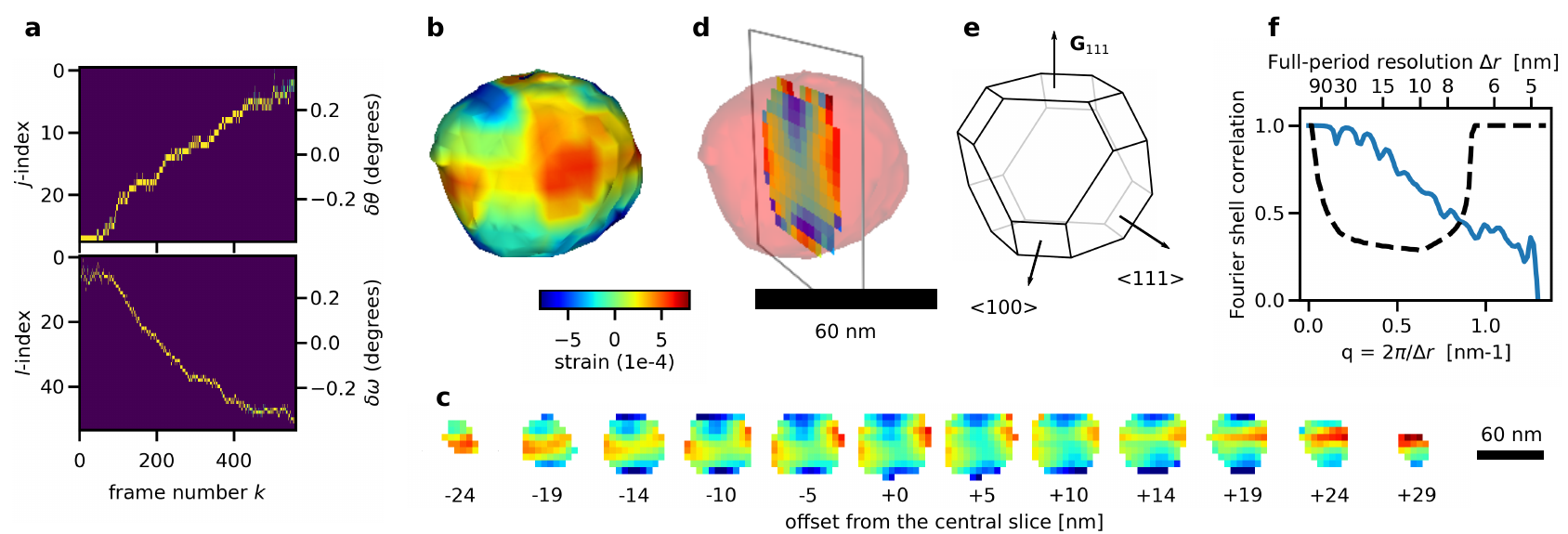}
            \caption{Experimental results for a selected particle. a) Assembly results presented as in Fig.~\ref{fig:sim}d. b) Amplitude isosurface colored according to the (111) strain component on the surface. c) and d) Cuts through the particle displaying the internal strain. e) Orientation of the reconstructed particle as compared to the expected shape. f) Fourier Shell Correlation analysis of the two half-datasets, estimating the resolution of the final reconstruction. The dashed line shows the half-bit criterion (eqn. 17 of ref.~\cite{vanHeel2005}). The criterion is corrected for the support by scaling the number of voxels in each shell by $(D/L)^2$, where $D$ and $L$ are the linear dimensions of support and phased volume, respectively~\cite{vanHeel2005}.}
            \label{fig:results}
        \end{figure*}
        
        Fig.~\ref{fig:sim} demonstrates the described assembly algorithm with a numerical example. The truncated-octahedral particle illustrated in Fig.~\ref{fig:intro}a was combined with an arbitrary phase, rendered from a spherical harmonic and a radial factor, and varying between -0.5 and 0.5 radians (Fig.~\ref{fig:sim}a). Arbitrary $(\theta,\omega)$ trajectories around the reciprocal lattice point $G_{111}$ were generated (Fig.~\ref{fig:sim}c), and diffraction patterns with applied Poisson noise generated (Fig.~\ref{fig:sim}b). The algorithm described above was then iteratively carried out for 100 iterations. Prior to assembly, the frames were pre-aligned along $\omega$ by applying the $\Omega_l$ operator such that their centers of mass lined up, saving computational effort by reducing the size of the $P_{jlk}$ matrix along $l$. Initial probabilities corresponding to a constant-velocity trajectory $\theta(k)$ and fixed $\omega(k)$ were used to generate the initial model according to Eq.~\eqref{eq:M}. A soft bias with $6\leq n_\sigma\leq 12$ was applied to $P_{jlk}$ as described above, in order to gently enforce a single-valued $\theta(k)$. A maximum particle dimension $D= 60$~nm constraint was applied, and the annealing parameter $\beta$ was initially set to $10^{-5}$ and then multiplied by $\sqrt{2}$ every 5 iterations. Fig.~\ref{fig:sim}d shows the angular probabilities from the assembly process, which recover the input trajectories very well (see Fig. S2 for residuals).
        
        The resulting diffraction volume model was phase-retrieved with the PyNX software~\cite{Favre-Nicolin2020} where 1000 reconstructions were carried out, each with a constant shrink-wrap support threshold across the range 0.1--0.5 times the average sample amplitude~\cite{Marchesini2003}. Each reconstruction consisted of 600 RAAR iterations~\cite{Luke2005} followed by 200 Error-Reduction cycles~\cite{Gerchberg1972,Fienup1982}. Before the support thresholding, the image was blurred using a Gaussian convolution kernel with $\sigma$ exponentially decreasing from 2 to 0.5 pixels from the first to the last RAAR step, following the PyNX default settings. The 20 reconstructions with the highest free log-likelihood metrics were selected and combined via mode decomposition, as described in ref.~\cite{Favre-Nicolin2020}. The reconstructed particle faithfully reproduces the three-dimensional shape of the input, with the features of the truncated-octahedral envelope clearly seen. The reconstructed volume also shows the main features of the original phases. These numerical results are analyzed further in Supplemental Figs. S3-S5.
        
        We now turn to the collection and analysis of experimental data. Shape-controlled Au nanoparticles of 60~nm diameter were synthesized, cleaned and supported in a conductive carbon matrix as described elsewhere~\cite{Erikson2014}. We have previously provided two-dimensional BCDI reconstructions of these model particles, and ref.~\cite{Bjorling2019} contains transmission electron micrographs of the sample. Experimental BCDI data were collected by placing a deposit of particle-carbon mixture on a Si$_3$N$_4$ membrane, and positioning the sample in the focused 10 keV X-ray beam. The X-ray focal spot (90~nm FWHM) is described in ref.~\cite{Bjorling2020}. The detector used was a Merlin Quad (Quantum Detectors) of pixel size \SI{55}{\micro\meter}, placed at $2\theta_{111}$, corresponding to the reciprocal lattice point $\mathbf{G}_{111}$ for Au, 320~mm from the sample. The sample deposit was stepped through the beam, collecting bursts of 1000 images at 100~Hz (8~ms exposure, 2~ms readout) at each position. No sample rotation was used. In total, 2900 such bursts were collected and analyzed, of which several hundred contained visible diffraction patterns. The bursts containing the brightest transient diffraction spots were selected for further analysis. Since the particles are uniform in size~\cite{Bjorling2019}, it can be assumed that the brightest rocking curves correspond to the most complete overlap of particle and beam.
        
        Several uncontrolled rocking curves were analyzed by diffraction volume assembly and phase retrieval as described above for the simulated data, with only minor tuning of assembly parameters. Fig.~\ref{fig:results} shows data from one selected particle. In panel (a), the recovered rotational trajectories $\theta(k)$ and $\omega(k)$ are plotted, showing fairly smooth paths with fluctuating velocities. The reconstructed three-dimensional particle shape is shown in panel (b) and displayed as individual slices in panel (c). A cut perpendicular to the diffraction plane is indicated in panel (d). The reconstructed shape closely resembles the shape expected for these truncated-octahedral particles. By comparing the reconstructed sample to its ideal counterpart (Supplemental Fig. S1) the $\varphi$ orientation can be determined, as shown for comparison in panel (e). The reconstructed voxel size is (4.5 nm)$^3$, and a particle is around 12-15 pixels across. At this pixel size and resolution, $\langle111\rangle$ and sometimes $\langle100\rangle$ facets can be discerned in the two-dimensional slices (Fig.~\ref{fig:results}c and S6-S14), whereas a smooth three-dimensional model (Figs.~\ref{fig:results}b) merely reveals the general shape.
        
        The colors in Fig.~\ref{fig:results} reflect the lattice strain component along the Bragg vector $\mathbf{G}_{111}$ (in the figure's vertical direction). This strain component is obtained from the reconstructed phase $\phi$,
        \begin{equation}
            \epsilon_{zz} = \frac{\partial u_z}{\partial z} = \frac{1}{|\mathbf{G}_{111}|} \frac{\partial \phi}{\partial z} ,
            \label{eq:strain}
        \end{equation}
        where $z$ denotes the vertical direction~\cite{Newton2010}. The results show strain patterns that are strongest at the particle surfaces, and which decay towards its interior. A lattice compression along $z$ is seen at the top and bottom surfaces, while the particle perimeter shows a lattice expansion in the same direction. In terms of surface strain, this is consistent with a slight lateral surface lattice expansion. Such an expansion would give a positive $\epsilon_{zz}$ component around the particle perimeter, and a negative contribution on the top and bottom as the lattice responds with a compressive elastic deformation normal to the surface.
        
        Reconstructions of a number of independent particles show similar three-dimensional strain patterns, as shown in Supplemental Figs. S6-S14. While this lends credence to the obtained results, individual particle reconstructions can also be validated by splitting one raw data set in two, then assembling and phase-retrieving each subset independently. This is routinely done in electron cryomicroscopy~\cite{vanHeel2005}, and has also been used for XFEL single-particle CDI~\cite{Ekeberg2015}. We apply this validation method here, with the modification that frames $K_k$ are split into even and odd $k$, as randomizing the subsets can produce too large gaps in the angular trajectories. The reconstructed subsets can then be compared using Fourier Shell Correlation (FSC)~\cite{Harauz1986} to obtain an estimate of the actual resolution, as shown in Fig.~\ref{fig:results}f for the selected particle. The half-bit information threshold for the current detector geometry and assembly conditions (dashed line) indicates an average three-dimensional resolution of around 8~nm~\cite{vanHeel2005,Ayyer2019}, which is sufficient to reliably map the observed strain fields.
        
        We have shown that it is possible to assemble a coherent diffraction volume from rocking curves with unknown angular trajectories. These volumes can be phase-retrieved, producing verifiable three-dimensional shape- and strain maps of crystalline nanoparticles. In general, this ability allows applying BCDI to new conditions not limited to beam-induced motions, including for example dynamic samples or particles in suspension, as long as the frame rate is fast compared to the rotation. For beam-induced rotations in particular, the method presented can significantly increase BCDI's robustness. In relaxing the tolerances on the rocking angle, the range of applicability of BCDI is extended to smaller and more technologically relevant particles which are prone to beam-induced sample rotation. By overcoming the rotation stability problems in high-flux experiments we therefore hope that this paper heralds the use of fourth generation synchrotrons for BCDI to their full potential.

    \begin{acknowledgments}
        ~\\
        Research conducted at MAX IV, a Swedish national user facility, is supported by the Swedish Research council under contract 2018-07152, the Swedish Governmental Agency for Innovation Systems under contract 2018-04969, and Formas under contract 2019-02496. This work has also received funding from the {\AA}Forsk Foundation (contract 17-408), the European Research Council (ERC) under the European Union’s Horizon 2020 research and innovation programme (contract 801847), from the Olle Engkvist foundation, from the Swedish Research council (contract 2018-00234), and from NanoLund.
    \end{acknowledgments}
    
    \appendix
    All code and raw data described in this article are deposited and available through ref.~\cite{cxidb-entry}.

%

\end{document}


\maketitle

	\begin{figure}[h]
  	\centering
		\includegraphics[width=.5\textwidth]{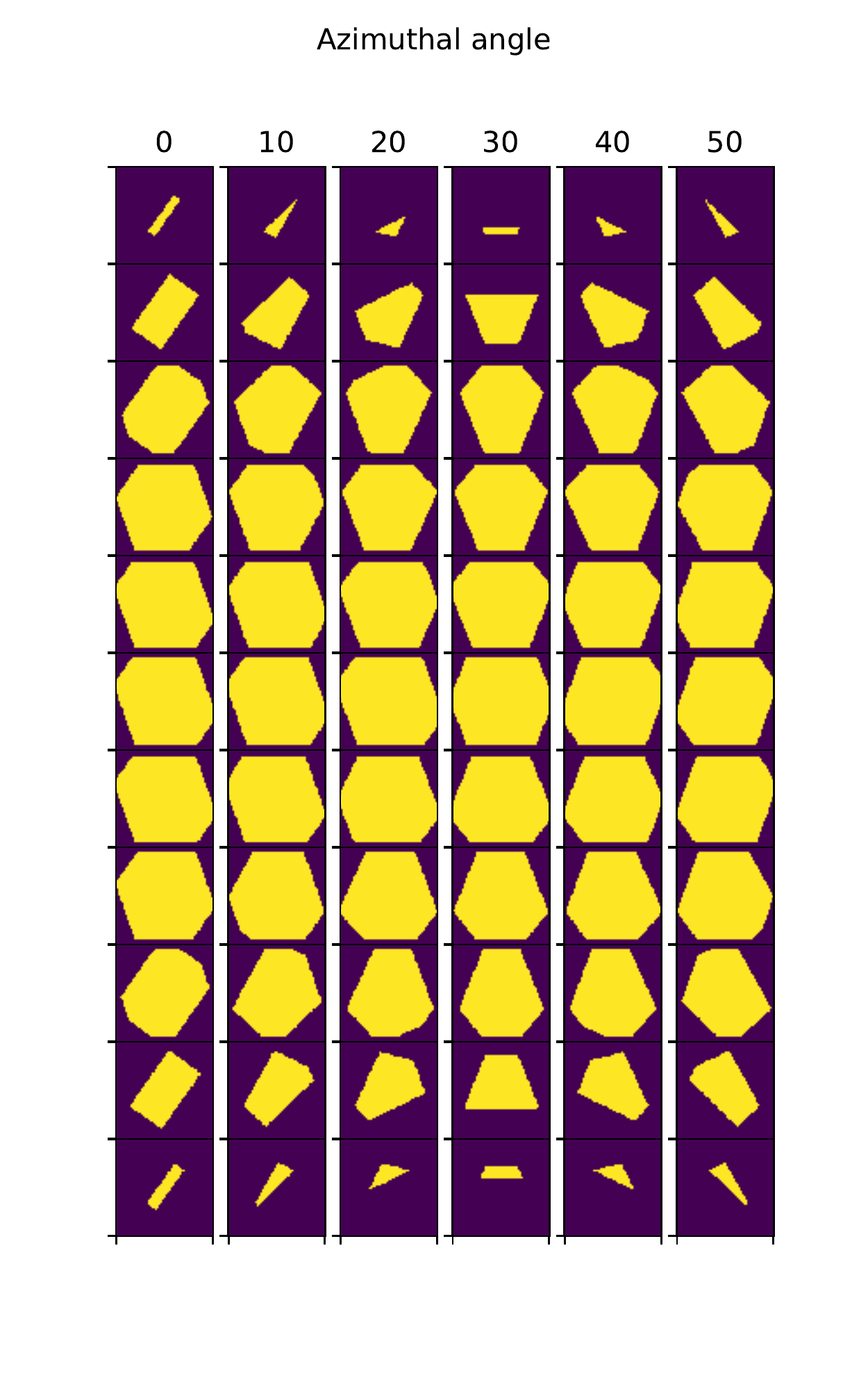}
		\caption{Slices through the model particle in Fig. 1, for comparison with the obtained reconstructions.}
	\end{figure}	
	
	\begin{figure}[h]
	\centering
	    \includegraphics[width=\textwidth]{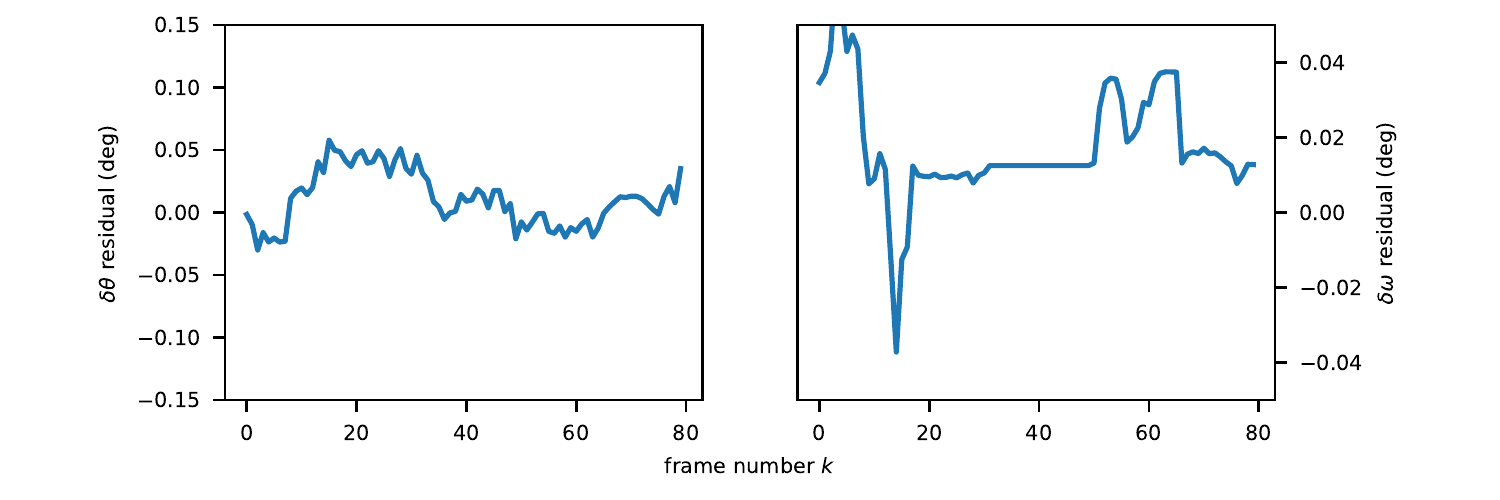}
		\caption{Residuals between the trajectory defined in the simulation (Fig. 2c) and the trajectory found in the volume assembly process (Fig. 2d). The residuals are calculated based on the centers of mass of the $P_{jk}$ and $P_{lk}$ matrices along $j$ and $l$, respectively.}
	\end{figure}
	\begin{figure}[h]
	
	\centering
	    \includegraphics[width=.8\textwidth]{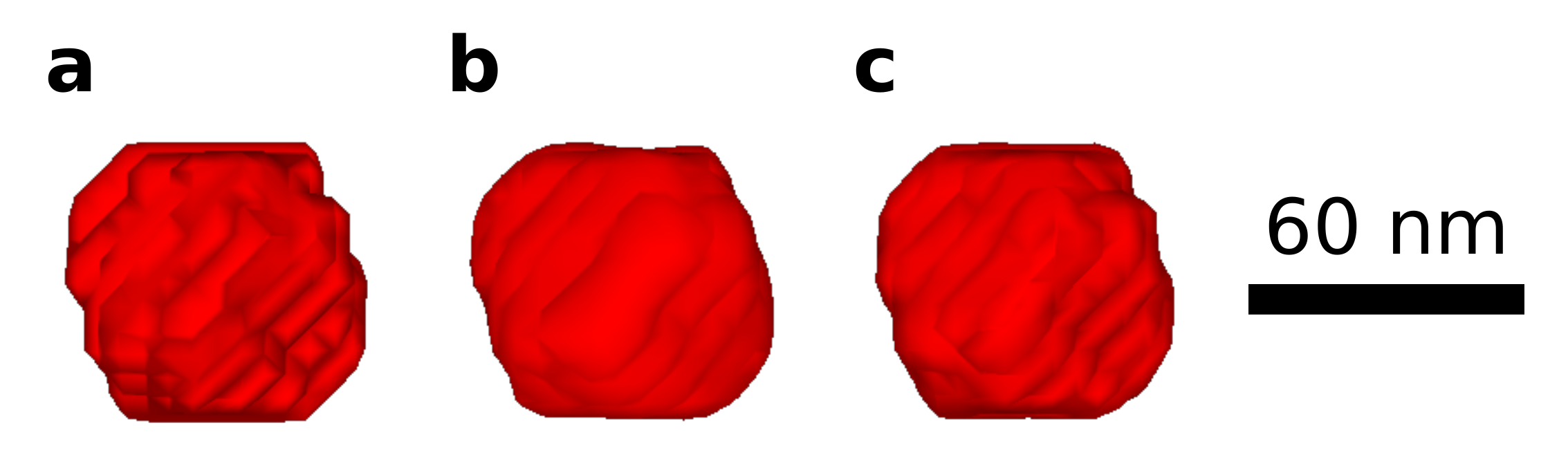}
		\caption{Three-dimensional isosurface representations of (a) the model particle used as input in the simulations (same as in Fig. 2a), (b) the output particle obtained after assembly and phase retrieval (same as in Fig. 2e). Panel (c) shows the particle obtained by simulating regularly spaced diffraction patterns and inverting these without assembly of unknown slices, with noise levels as in Fig. 2b.}
	\end{figure}
	
    \clearpage
	
	\begin{figure}[h]
	\centering
	    \includegraphics[width=\textwidth]{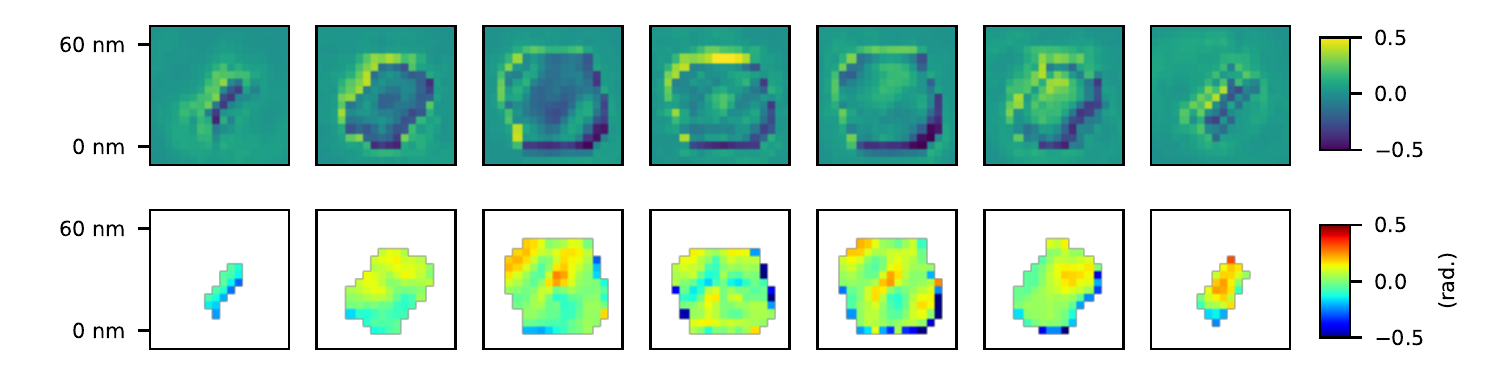}
		\caption{Difference maps between the "ground truth" input particle (Fig. 2a) and the output particle obtained from volume assembly and phase retrieval (Fig. 2e). The top row shows amplitude differences, while the bottom row shows phase differences taken after positional alignment, phase ramp removal, and multiplication by arbitrary phase factors. The output particle reproduces the shape of the input particle. Differences in phase are seen, but they are comparable to those caused by the pure phase retrieval process (see next figure).}
	\end{figure}
	
		\begin{figure}[h]
	\centering
	    \includegraphics[width=\textwidth]{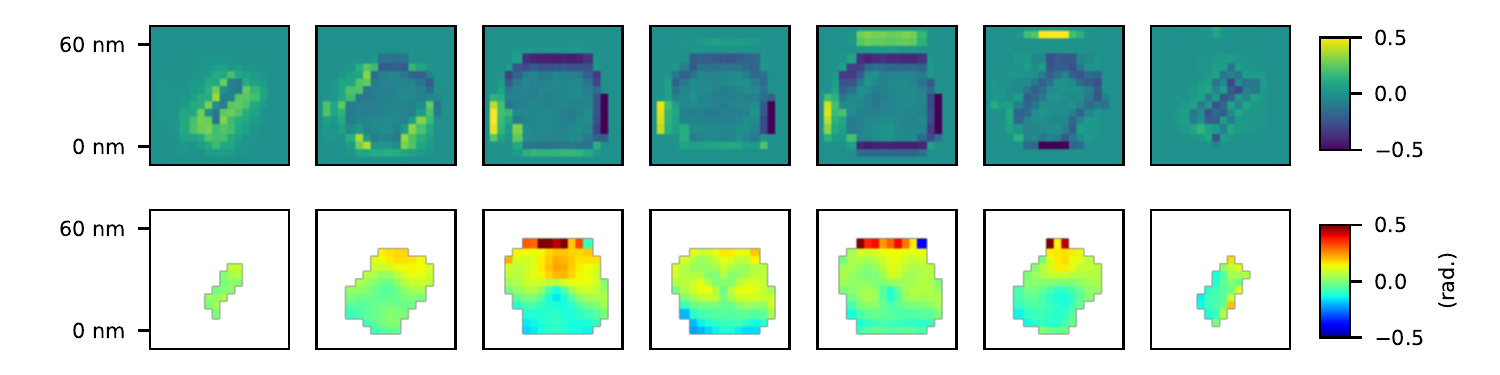}
		\caption{Difference maps between the "ground truth" input particle (Fig. 2a) and a particle obtained by phase retrieval of perfectly assembled, regularly spaced diffraction patterns. The phase differences come from the application of shot noise (shown in Fig. 2b), phasing artefacts, and imperfect phase alignment and phase ramp removal. While slightly smaller, the phase differences observed are still comparable in magnitude to those in Fig. S4 above. This means that scrambling the angular trajectory and reassembling the diffraction volume doesn't dramatically degrade the reconstruction.}
	\end{figure}
	
	\clearpage
	
	\begin{figure}
	  \centering
		\includegraphics[width=\textwidth]{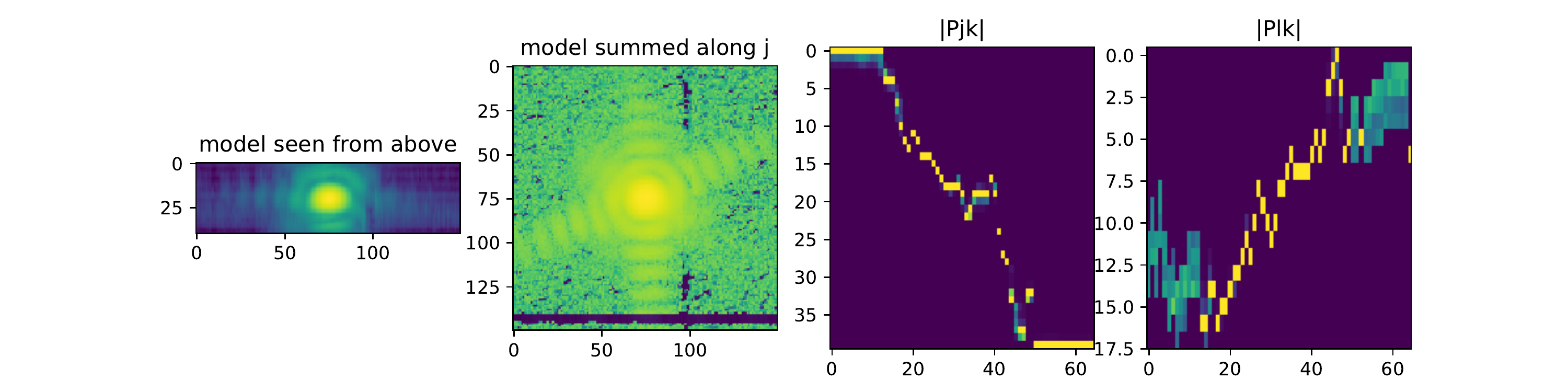}
		\includegraphics[width=\textwidth]{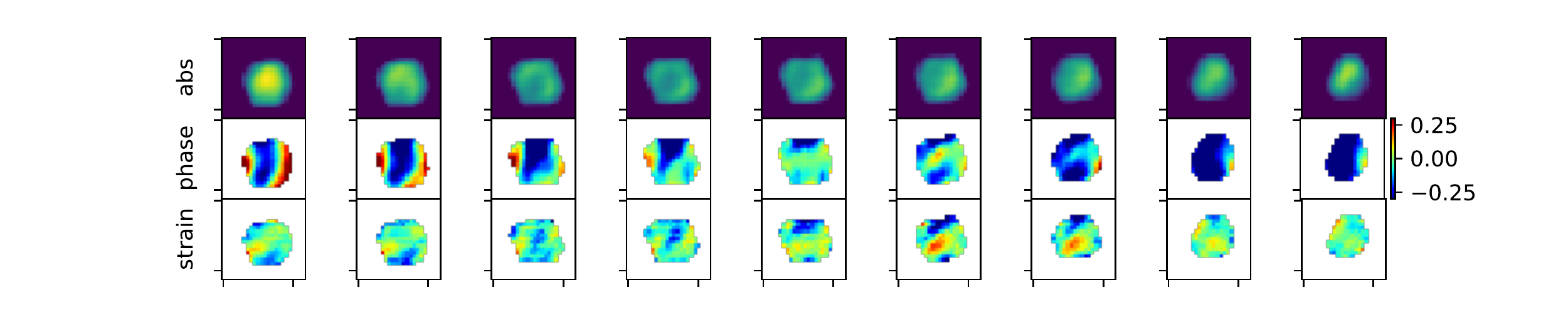}
		\caption{Particle \#101\_129. For this particular hit, one half-dataset did not assemble, and there is therefore no FSC validation.}
	\end{figure}	
	
	\begin{figure}
	  \centering
		\includegraphics[width=\textwidth]{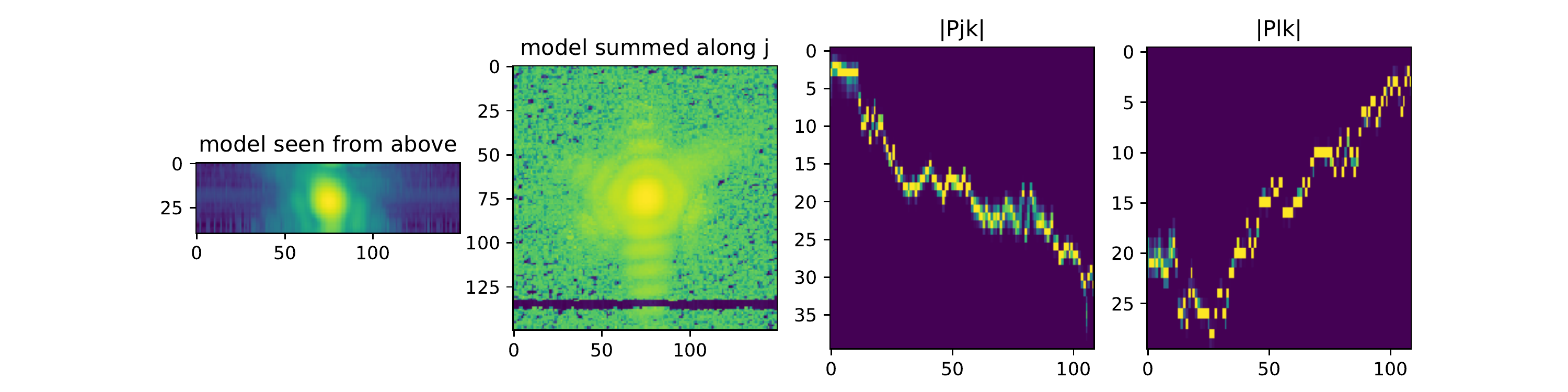}
		\includegraphics[width=\textwidth]{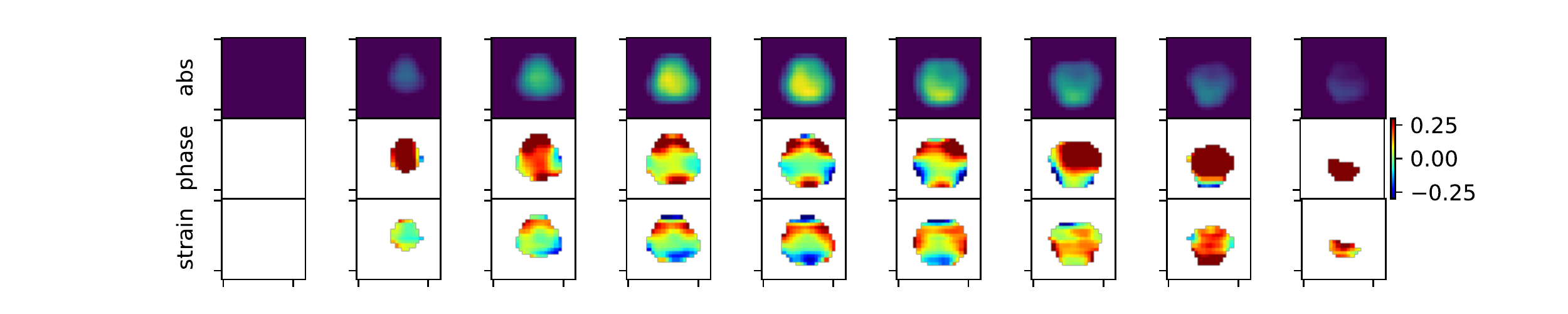}
		\includegraphics[width=.6\textwidth]{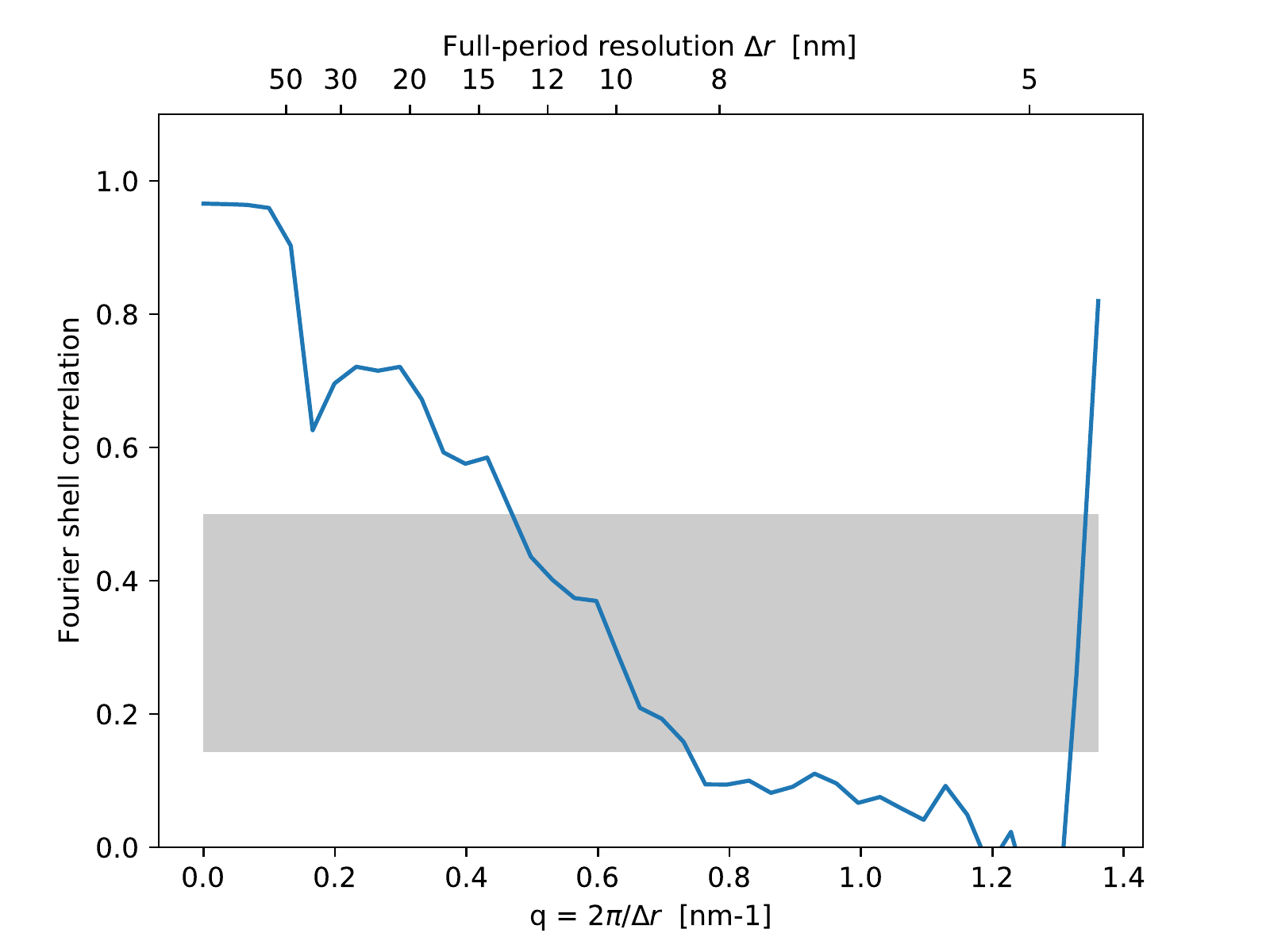}
		\caption{Particle \#102\_109}
	\end{figure}	
		
	\begin{figure}
  	\centering
		\includegraphics[width=\textwidth]{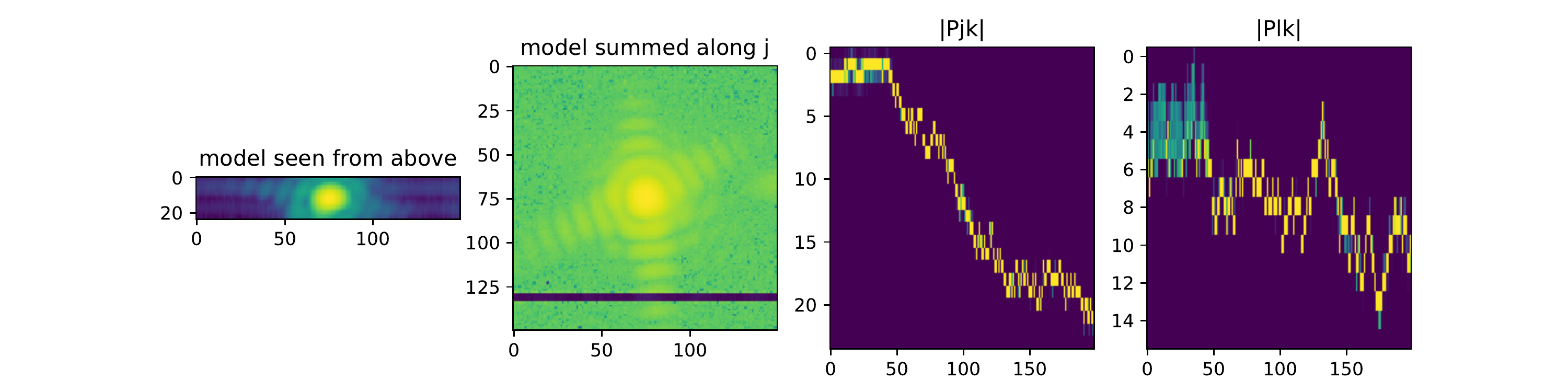}
		\includegraphics[width=\textwidth]{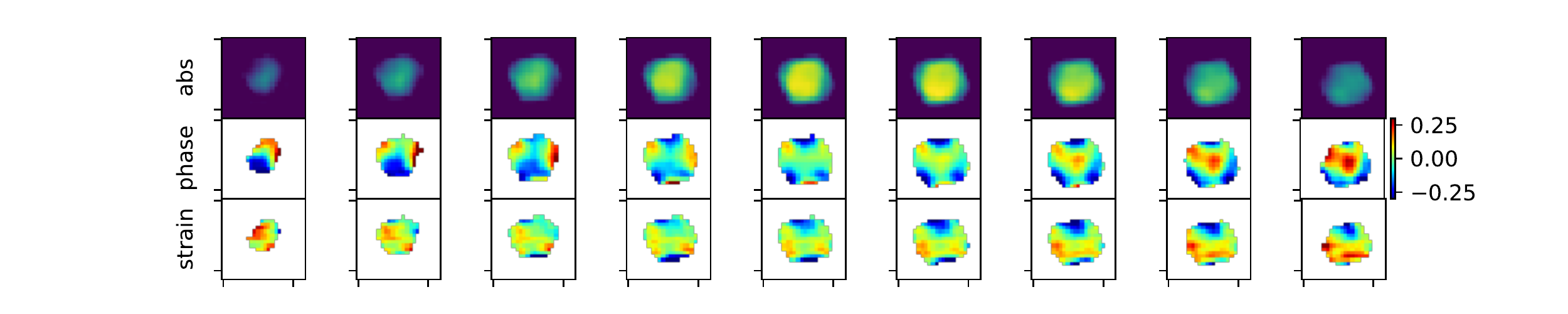}
		\includegraphics[width=.6\textwidth]{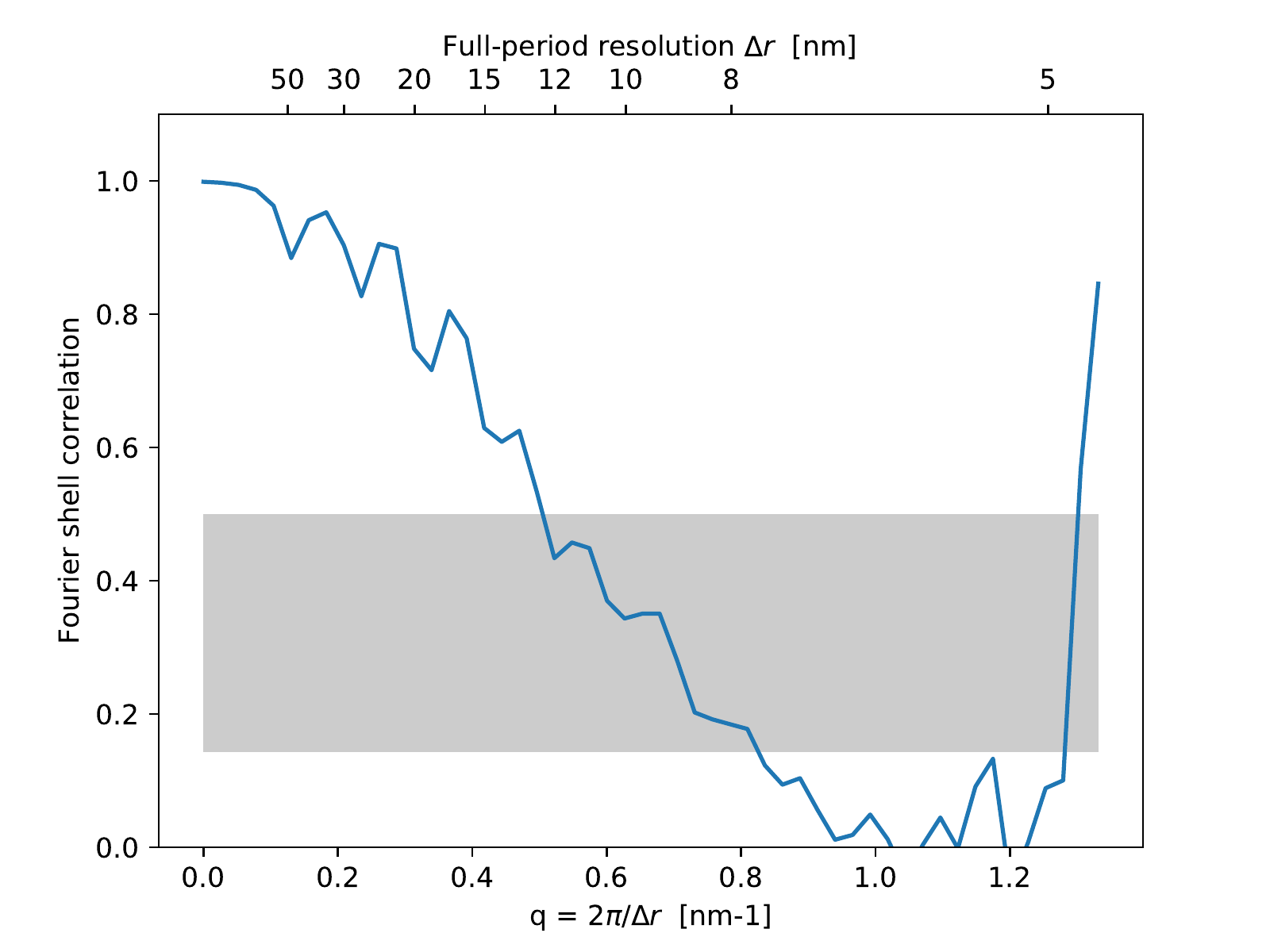}
		\caption{Particle \#111\_207}
	\end{figure}	
		
	\begin{figure}
	  \centering
		\includegraphics[width=\textwidth]{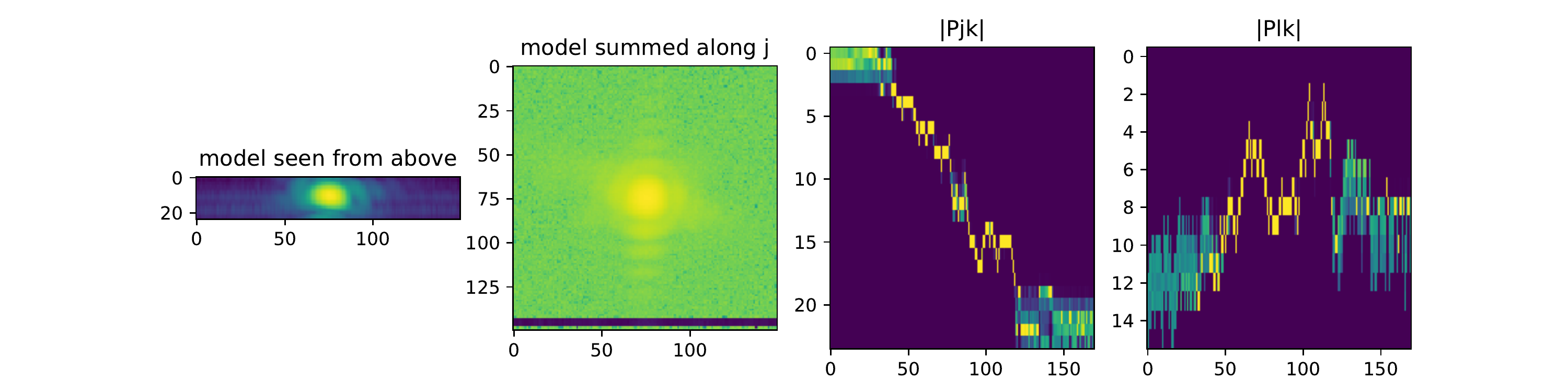}
		\includegraphics[width=\textwidth]{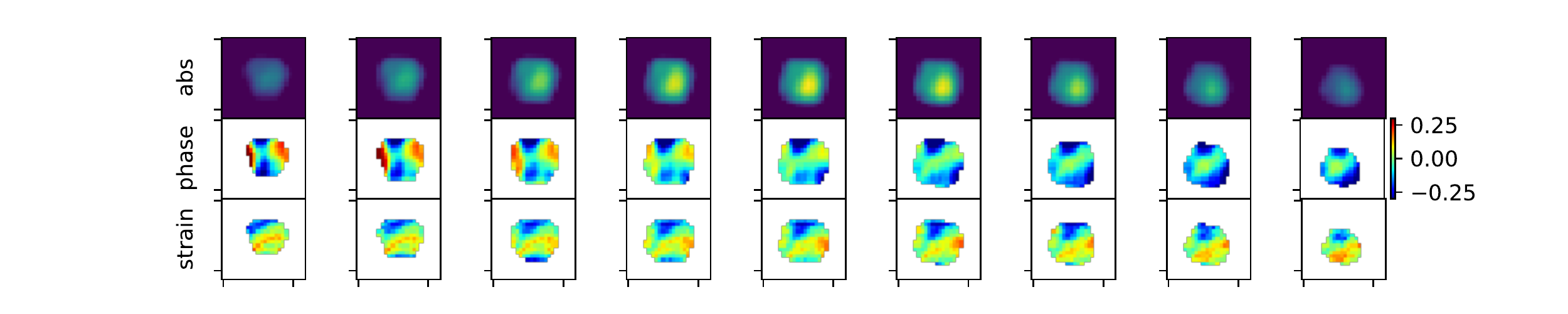}
		\includegraphics[width=.6\textwidth]{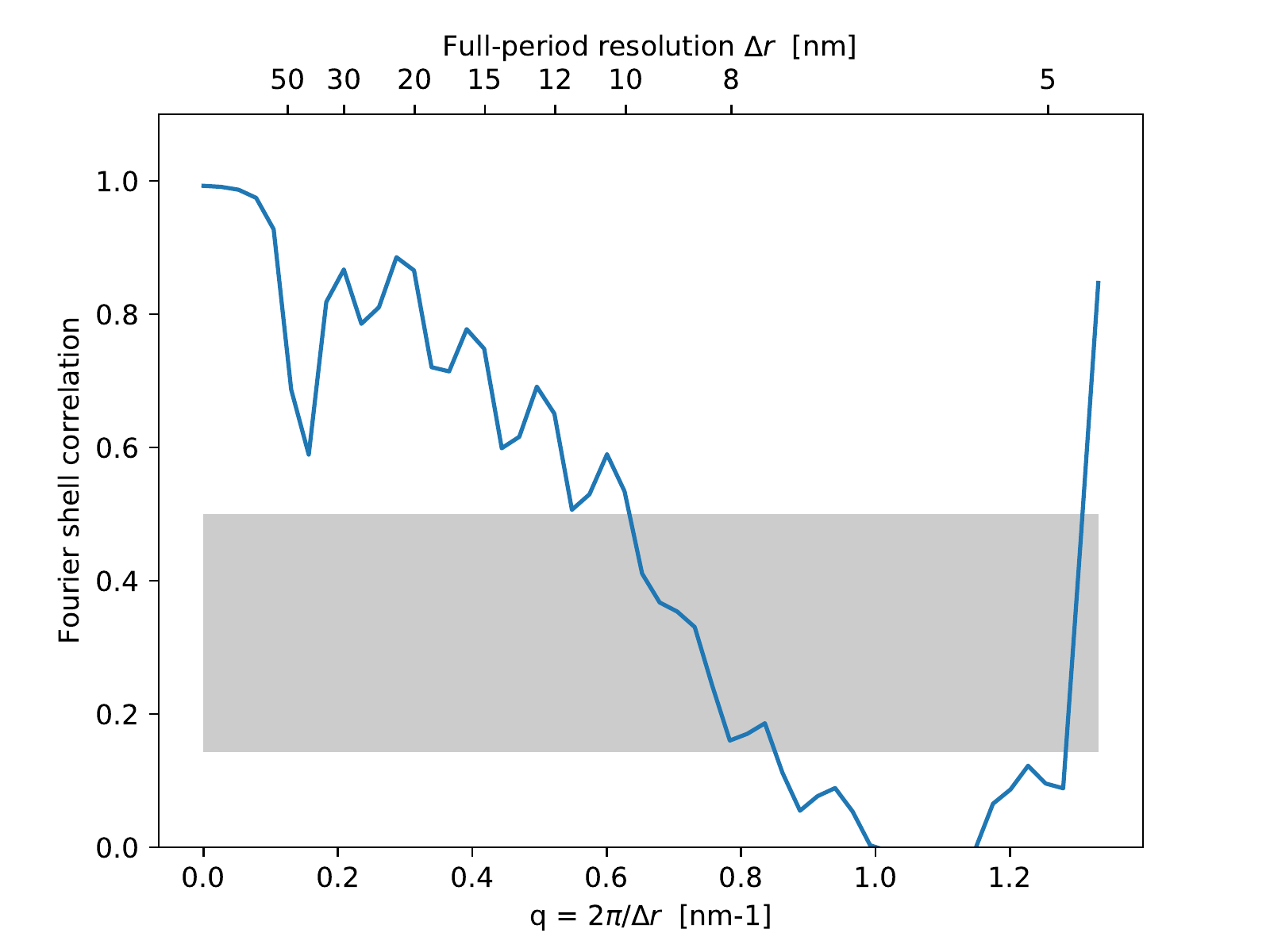}
		\caption{Particle \#111\_042}
	\end{figure}	
		
	\begin{figure}
	  \centering
		\includegraphics[width=\textwidth]{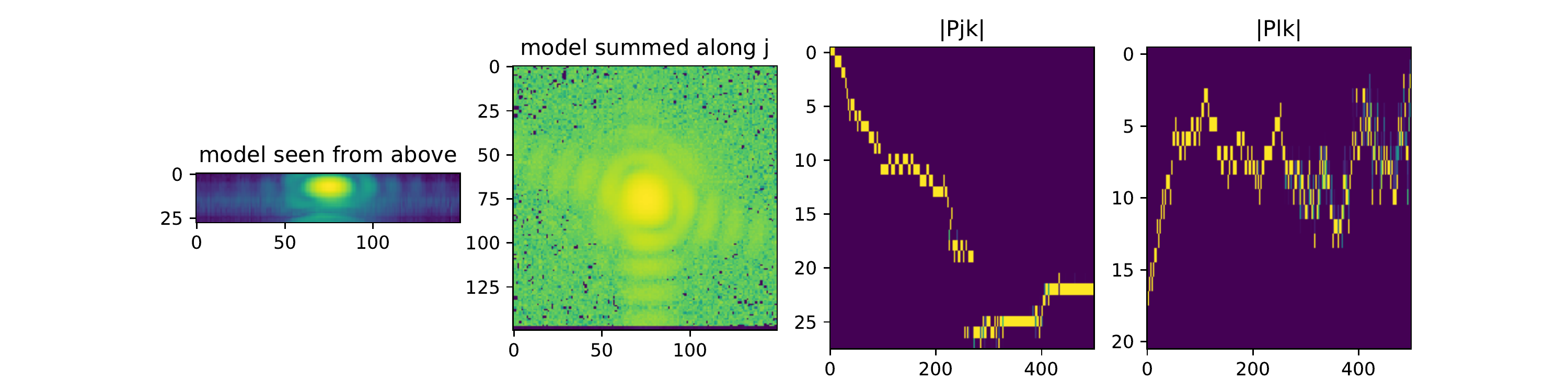}
		\includegraphics[width=\textwidth]{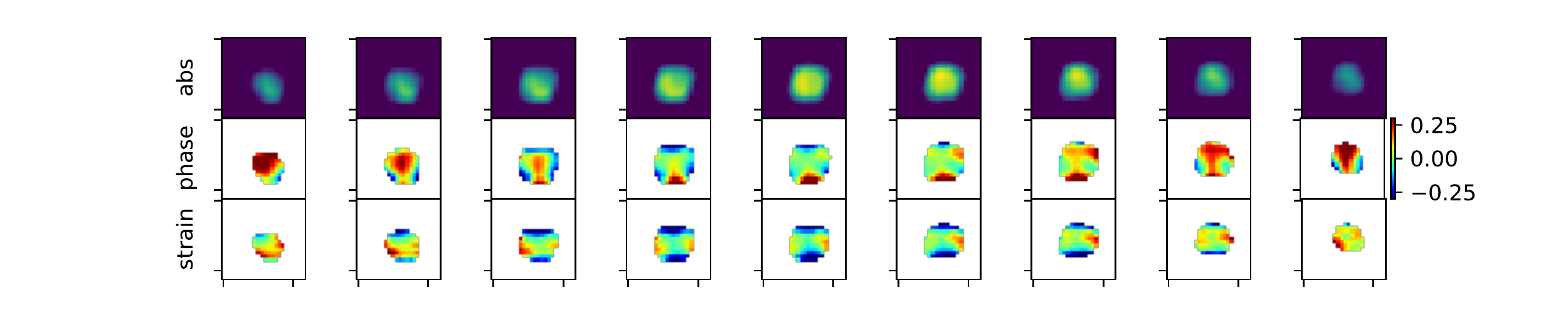}
		\includegraphics[width=.6\textwidth]{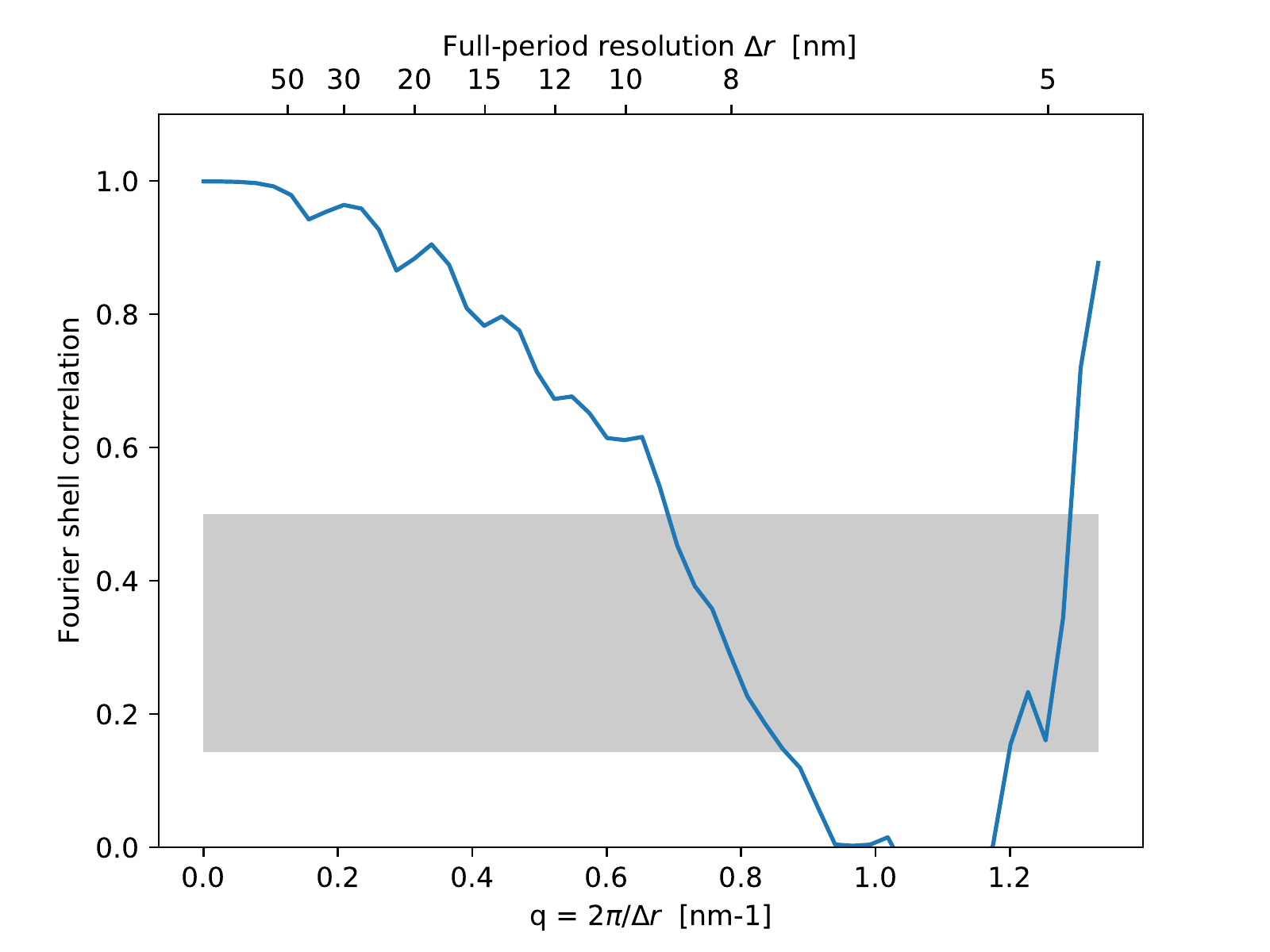}
		\caption{Particle \#19\_71}
	\end{figure}	
		
	\begin{figure}
	  \centering
		\includegraphics[width=\textwidth]{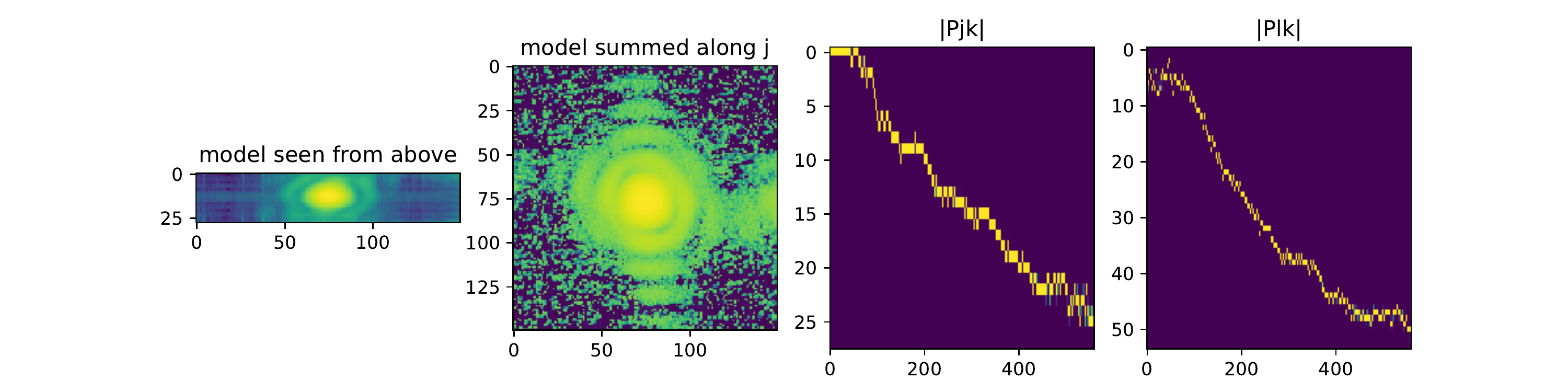}
		\includegraphics[width=\textwidth]{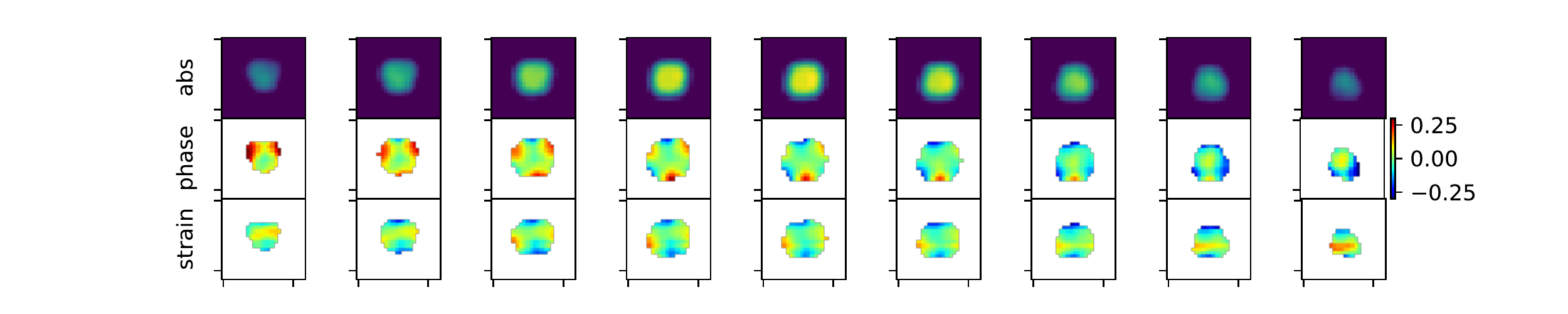}
		\includegraphics[width=.6\textwidth]{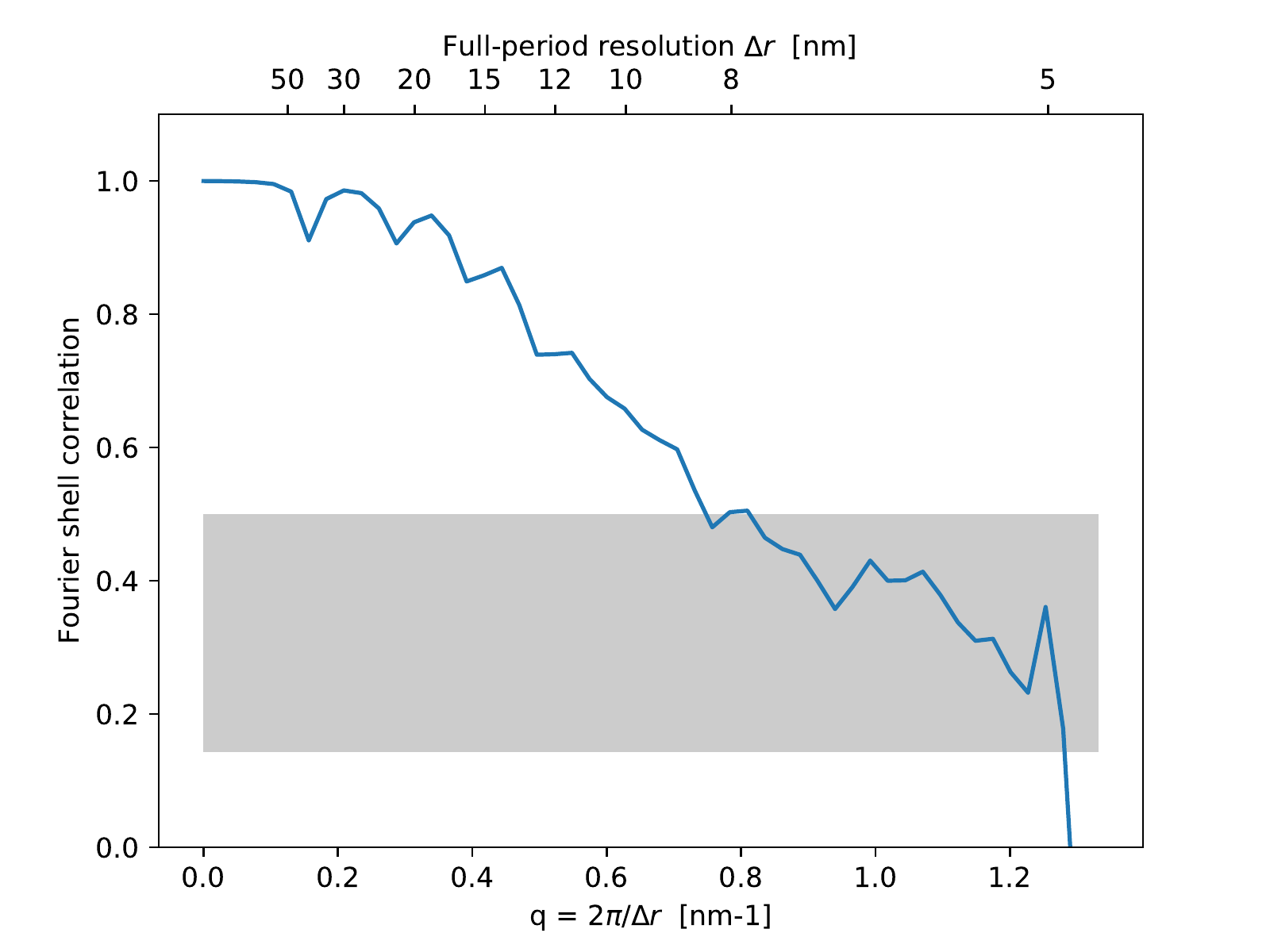}
		\caption{Particle \#15\_569}
	\end{figure}	
		
	\begin{figure}
	  \centering
		\includegraphics[width=\textwidth]{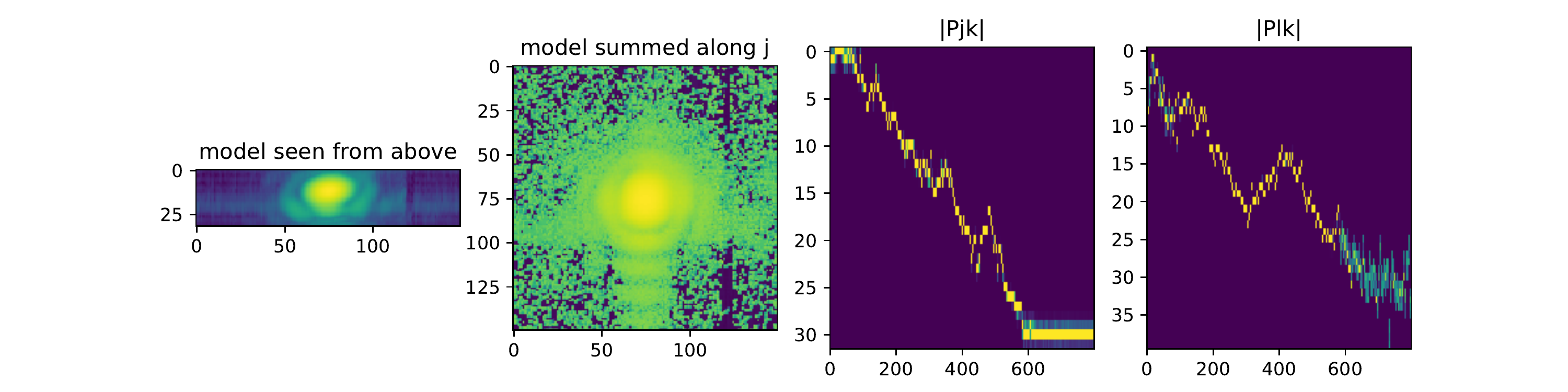}
		\includegraphics[width=\textwidth]{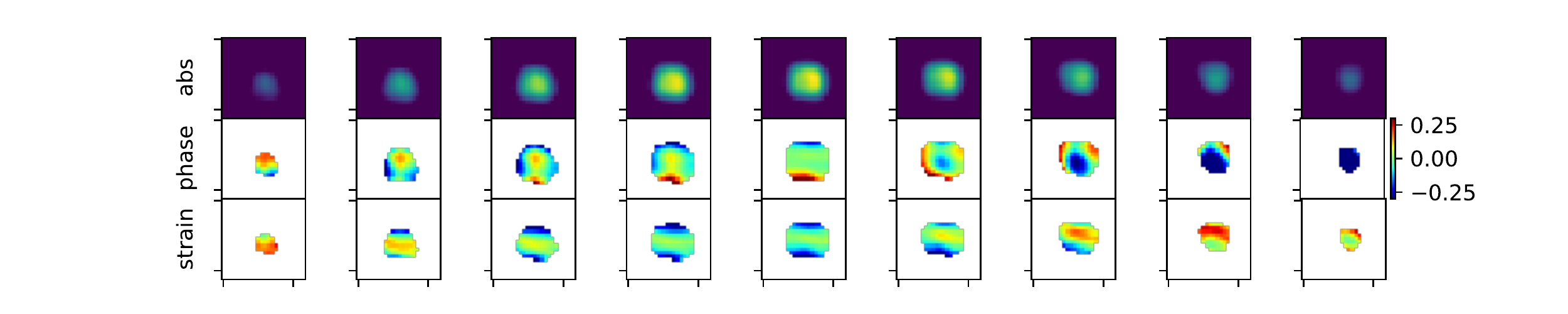}
		\includegraphics[width=.6\textwidth]{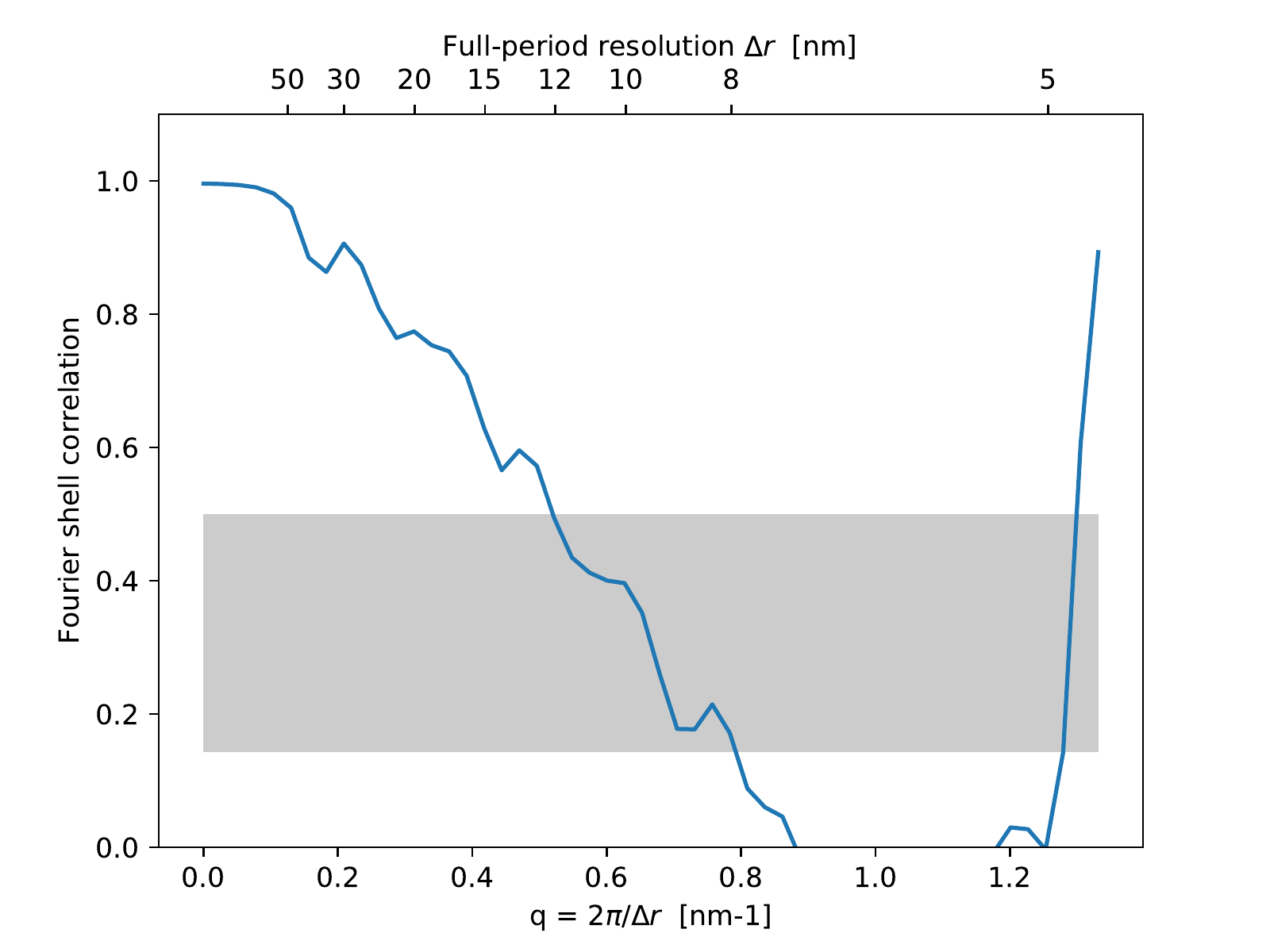}
		\caption{Particle \#16\_57}
	\end{figure}	
		
	\begin{figure}
	  \centering
		\includegraphics[width=\textwidth]{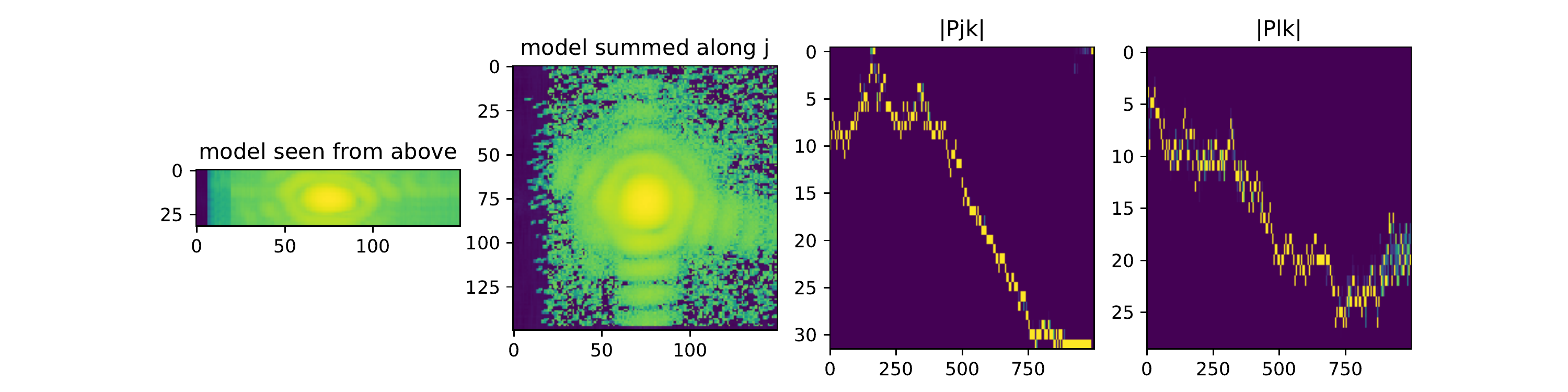}
		\includegraphics[width=\textwidth]{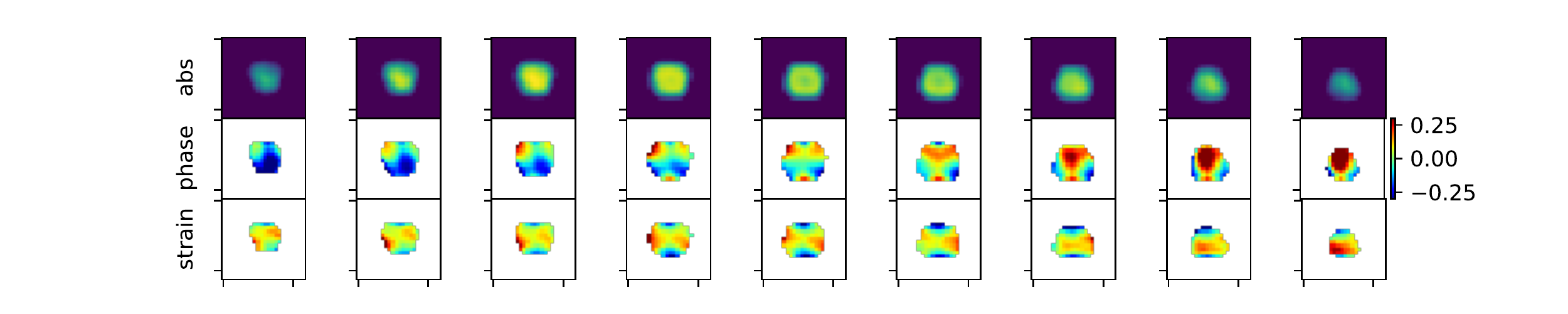}
		\includegraphics[width=.6\textwidth]{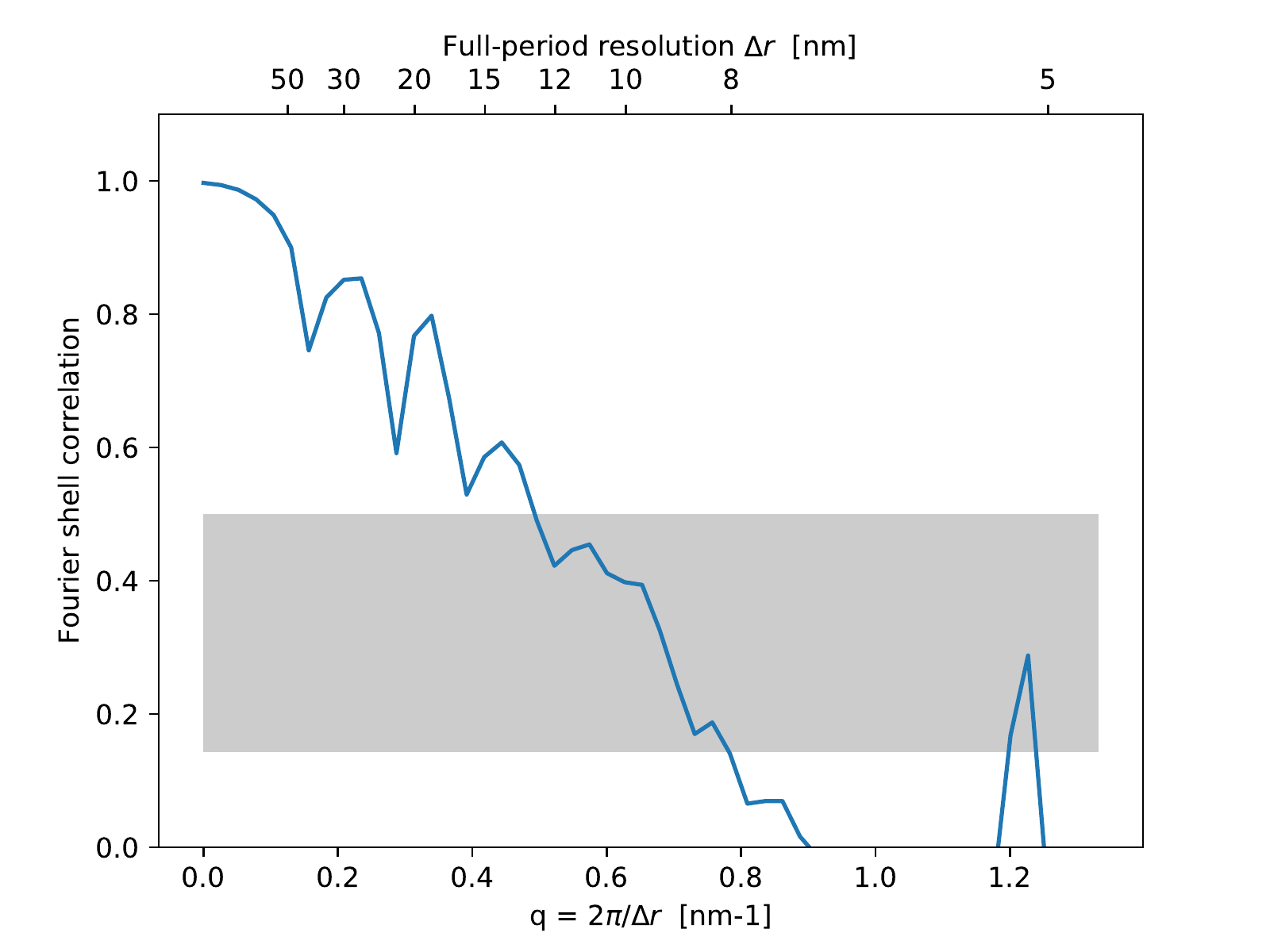}
		\caption{Particle \#17\_8}
	\end{figure}	
		
	\begin{figure}
	  \centering
		\includegraphics[width=\textwidth]{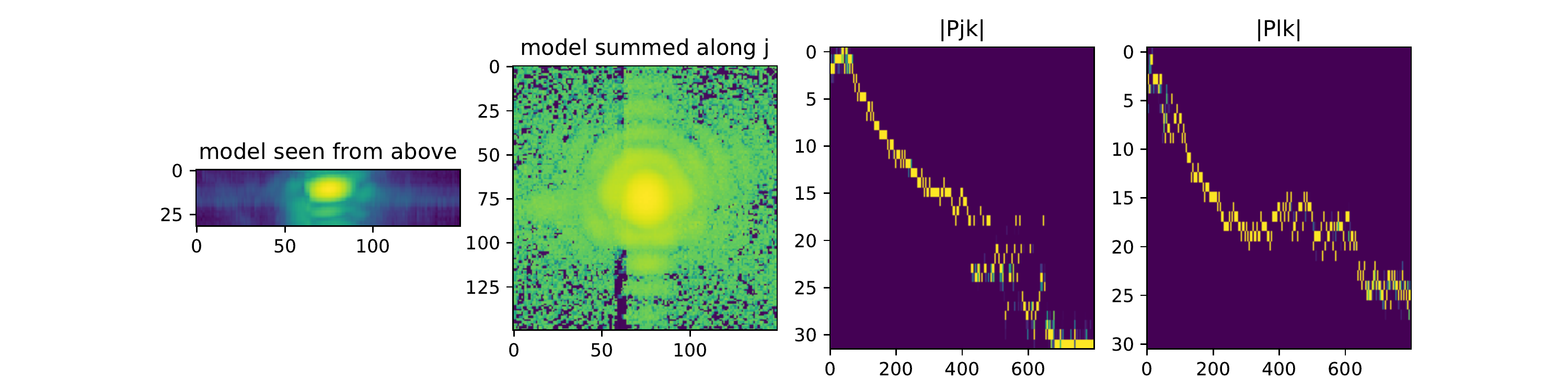}
		\includegraphics[width=\textwidth]{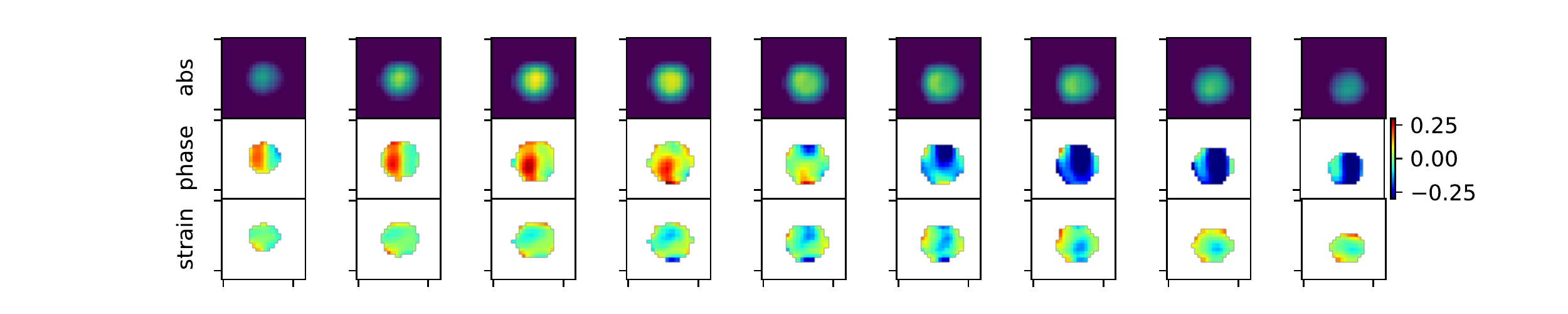}
		\includegraphics[width=.6\textwidth]{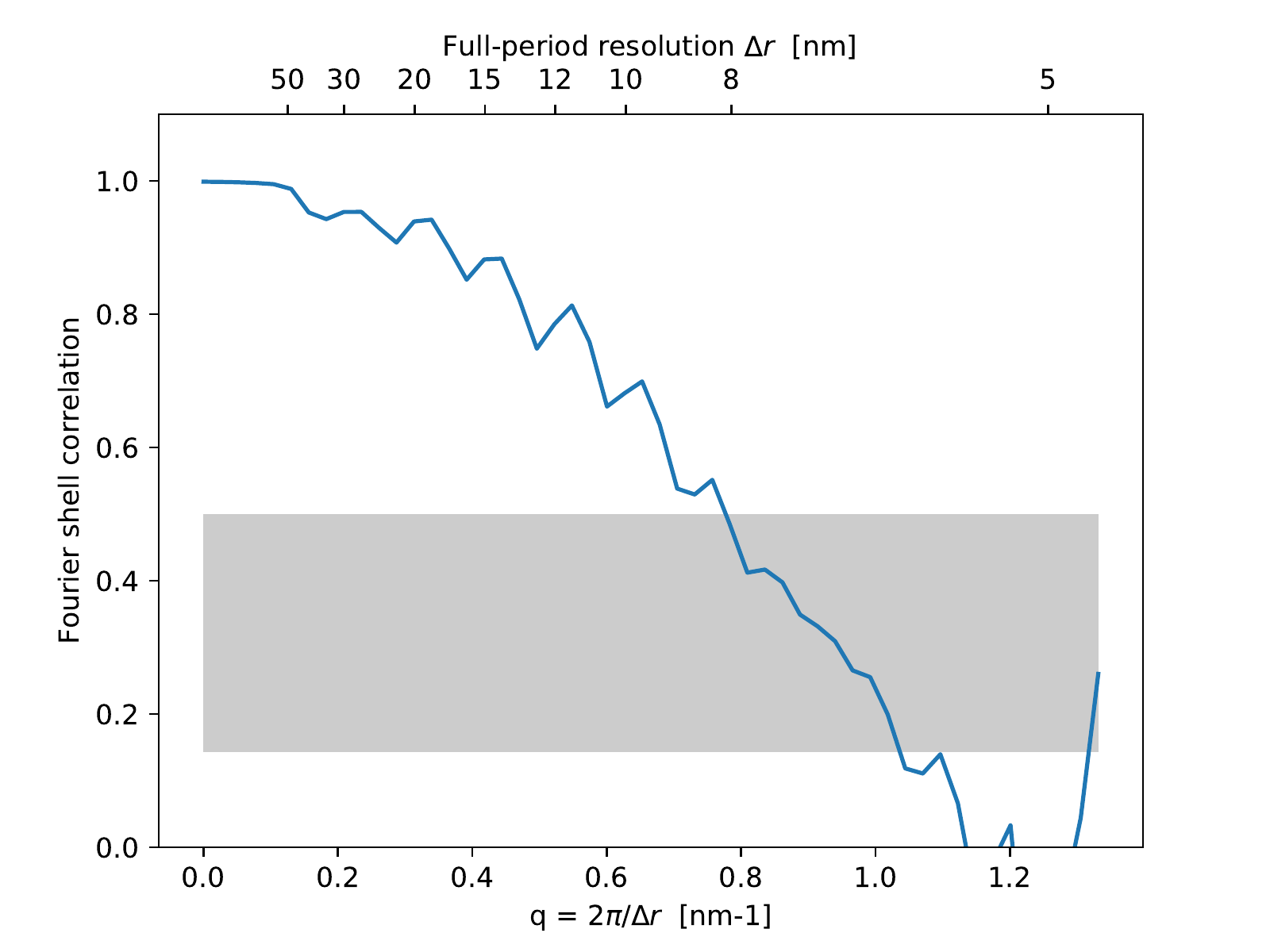}
		\caption{Particle \#21\_23}
  \end{figure}